%% file: groupc2004.tex
\documentclass[12pt]{article}
 
\usepackage{epsfig,rotating}
\usepackage{cite}

\newcommand{\ksp}{K^{0}_{S}\> p}
\newcommand{\kspb}{K^{0}_{S}\> \bar{p}}
\newcommand{\ksppb}{K^{0}_{S}\> p\>(\bar{p})}
\newcommand{\coll}{Collaboration}
\newcommand{\etal}{ {\it et al.,} }

\newcommand{\mev}{\; \mathrm{MeV}}

\def\ep{\epsilon}

\def\beq{\begin{equation}}
\def\eeq{\end{equation}}
\def\beeq{\begin{eqnarray}}
\def\eeeq{\end{eqnarray}}

\def\bom#1{{\mbox{\boldmath $#1$}}}
\def\to{\rightarrow}

\def\nn{\nonumber}

\def\ID{1 \kern -.45 em 1}
\def\RS{{\scriptscriptstyle\rm R\!.S\!.}}

\def\sp{{\bom {Sp}}}

\begin{document}

\clearpage
\pagestyle{empty}
\setcounter{footnote}{0}\setcounter{page}{0}%
\thispagestyle{empty}\pagestyle{plain}\pagenumbering{arabic}%

\hfill CERN-PH-TH/2004-184 

\hfill ANL-HEP-CP-04-90

\hfill September 2004 
 
\vspace{2.0cm}
 
\begin{center}
 
\vskip 0.8in plus 2in
 
{\Large\bf   Hadronic Final States }
 
\vspace{3.0cm}

{\large S.~Chekanov $^a$, 
M.~Dasgupta  $^b$, 
D.~Milstead $^c$}
 
{\begin{itemize}
\itemsep=-1mm
 
\normalsize
\item[$^a$]

\small
HEP division, Argonne National Laboratory,
9700 S.Cass Avenue,  
Argonne, \\ IL 60439
USA \\ E-mail: chekanov@mail.desy.de

\normalsize
\item[$^b$]
 
\small
Theory Division, CERN  
CH-1211, Geneva 23, Switzerland \\
E-mail: mrinal.dasgupta@cern.ch 
 
\normalsize
\item[$^c$]
 
\small
Stockholm University, Fysikum, 10691 Stockholm, Sweden \\
E-mail: milstead@physto.se

\end{itemize}
}
 
\normalsize
\vspace{2.0cm}
 
\begin{abstract}
A summary is given of the work which was presented in the working 
group C "Hadronic Final States" at the DIS04 workshop (Strbsk\'e Pleso,~Slovakia~, 
April 14-18, 2004). 
Progress in experimental tests and the 
theoretical development of perturbative and non-perturbative QCD is reported. 
\end{abstract}
 
\end{center}

\newpage
\setcounter{page}{1}

\input{intro}
\input{jetshigh}

\input{jetshighth}

\section{Event shape variables}
A class of observables that are perhaps best suited for testing 
QCD are event shape variables (for a recent review see \cite{dassalrev}). 
While the infrared and collinear safety of these observables allows one to 
make perturbative predictions, event shape distributions receive infrared and 
collinear enhancements, in the form of large logarithms, that require resummation. 
Resummation tools and predictions 
in turn follow from a sound understanding of QCD dynamics in the most important kinematic regions.

However, before one can confront resummed perturbative 
predictions with the accurate experimental data available for several event shapes, one usually has to account for power corrections. 
These are non-perturbative effects that, for most event shapes, scale as $1/Q$, 
$Q$ being the hard scale involved in the reaction, and make a significant contribution to many event shape observables. 
Over the past decade this problematic aspect of event shapes has been turned into an advantage since theoretical ideas now exist 
that permit a description of power corrections based on renormalons (for a review see \cite{Beneke}). 
The sizable power corrections involved  in event-shape studies make this a valuable testing 
ground for ideas on non-perturbative dynamics. Hence, event shapes allow one to test 
at once both perturbative and non-perturbative aspects of QCD.

\begin{figure}[!thb]
\begin{center}
\centerline{\epsfxsize=3.5in\epsfbox{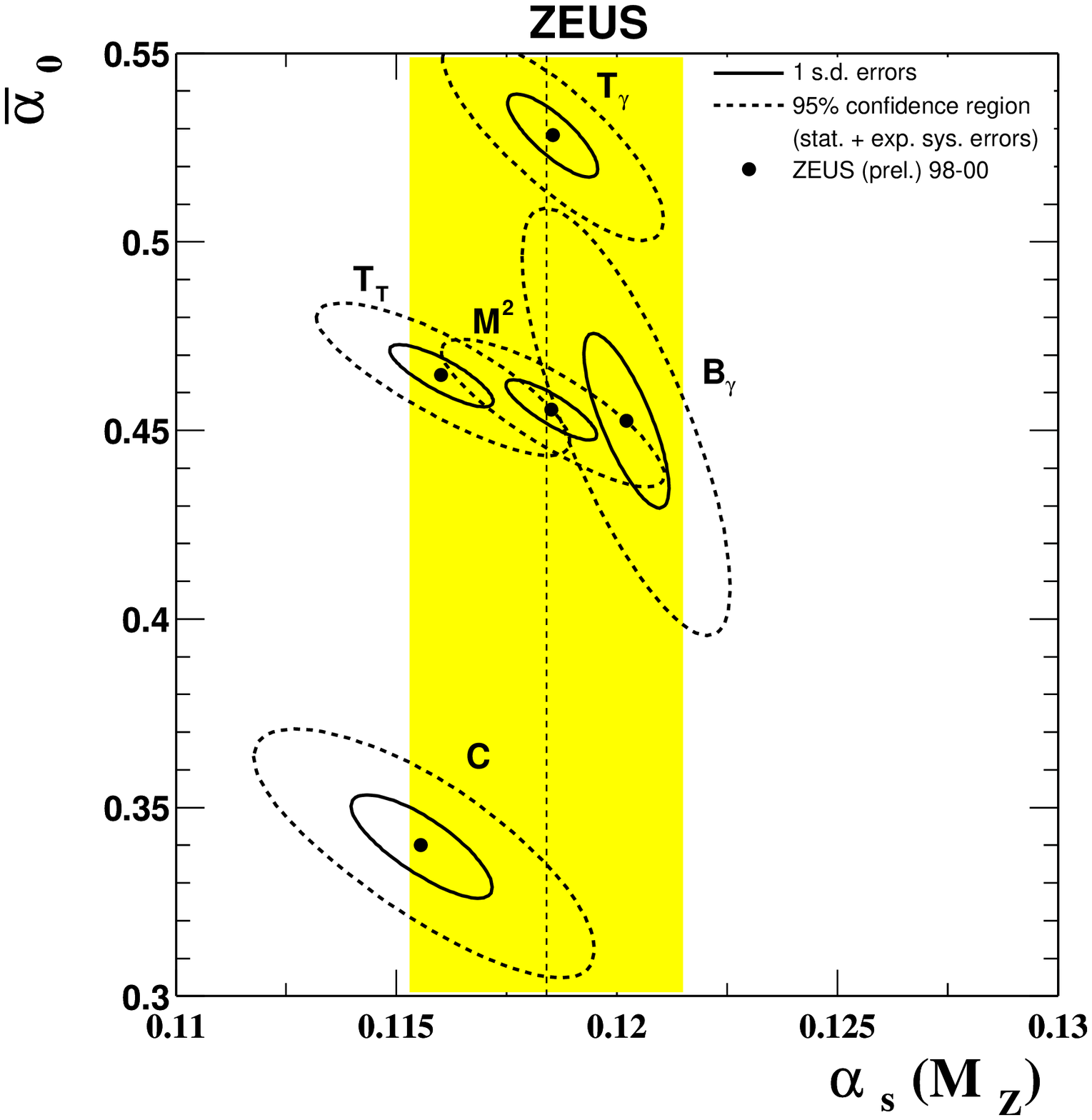}}
\caption[*]{Fitted values of  $\bar{\alpha_0}$ and $\alpha_s$ from 
event shape variable distributions, as measured by ZEUS.
\label{figz}
}
\end{center}
\end{figure}

Both the ZEUS \cite{zev} and H1 \cite{h1ev} experiments showed new results of 
event shape distributions in DIS, measured in the 
current region of the Breit frame.  Fig.~\ref{figz} shows the results of 
fits to event shape variable 
distributions, of resummed Next-to-leading logarithm (NLL) calculations matched to NLO calculations, with a power correction
ansatz for the hadronisation step. The extracted values of 
$\bar{\alpha}_0$ and $\alpha_s$ are shown for a different 
range of event shape variables. Good consistency for 
the different variables is obtained except for the $C$ variable. For this variable, there is much sensitivity to the 
choice of the fit range.  

At this meeting there was a presentation on resumming 
Sudakov logarithms and power corrections for event shapes within the framework of dressed (i.e fake massive) gluon exponentiation (DGE) 
\cite{Gardiraths,Lorenzo}.
The basic idea of DGE is to use a massive gluon as the kernel 
for exponentiation, with the gluon mass effectively 
accounting for resummation of renormalon bubbles diagrams on each gluon:
\begin{equation}
S(\nu,Q^2) = \int_0^\infty d v\left( e^{- \nu
  v} - 1 \right) \left( \left. \frac{1}{\sigma_{\mathrm{tot}}} 
  \frac{d \sigma}{d v} \right|_{\mathrm{SDG}} \right)
  . 
\end{equation}
Here $v$ is the event shape variable and $\nu$ is a Mellin variable conjugate to it while the suffix ${\mathrm{SDG}}$ denotes 
the single dressed gluon result. 
The above equation implies that the distribution for the emission 
of multiple independent dressed gluons can be obtained by 'exponentiating' 
the single dressed gluon result on the RHS of the above (in other words taking the inverse Mellin transform of $e^
{S(\nu)}$).
 This result preserves 
all the information present in the resummed perturbative 
next-to-leading log (NLL) results that are known for several observables. It also resums \footnote{In the leading power of $\beta_0$.} 
(due to the dressed gluon), certain sub-leading logarithms whose 
coefficients would grow as $n!$, for large n, 
at $n^{{\mathrm{th}}}$ perturbative order . Additionally on Borel transforming the running coupling $\alpha_s (k)$ in the expression for $S$, 
one generates singularities in the Borel plane, 
whose residues correspond to power corrections.

This is contrast to the Dokshitzer-Webber (DW) \cite{DokWeb} 
approach in which one 
exponentiates the single massless gluon result and introduces a universal 
infrared finite extension of $\alpha_s$ to regulate the Landau singularity. 
In the DW approach the power correction then emerges as a shift of the perturbative distribution by an amount proportional to $1/Q$. While the shift has been known to be an excellent practical approximation for a wide range of values of $v$, it is well known that in the region of very small $v$, $v \sim \Lambda/Q$, the shift approximation breaks down due to increasing importance of higher-orders in the series of non perturbative terms $\Lambda^n/v^n Q^n$. The shift must then be replaced by a smearing of the perturbative spectrum with a shape function \cite{KS}. The dressed gluon exponentiation mimicks the shape function approach and,  in fact, 
can be considered as a renormalon model for the full shape function. 
Phenomenological applications based on DGE can be found in \cite{Gardiraths}.
For most recent results on the $C$ parameter and a new class of event shapes (angularities) see \cite{Lorenzo}.

It should be emphasised that the DGE, while attempting to maximise 
information obtained from renormalon resummed 
perturbation theory, is once again only a model.
In particular it is well known that use of a massive gluon is not strictly appropriate, being too inclusive an approximation, for the case of event shapes. The DGE fails at present to address the issue of gluon decay, which is included for instance in the shift picture. However it is interesting to note that renormalons included via DGE can reproduce the basic features of the series of power corrections resummed into non-perturbative shape functions.

Regarding purely perturbative resummed results there have been significant 
developments in NLL resummations for observables such as event shapes.
Results were presented, at this conference, on automated 
resummations for QCD final state observables \cite{salamproc}.
The automation of resummation methods for generic observables in 
different processes is a big step forward since resummations done on a case-by--case basis were tedious and often error prone. This was in contrast to fixed order computations for event shapes where Monte Carlo programs 
exist that generate 
NLO results for several 
different observables without any laborious calculations 
required on the part of the user. 

The approach encoded in the program CAESAR is a hybrid analytical and numerical method. The first step is to obtain the parametric dependence of the 
observable on final state soft and collinear 
particle momenta. This must be (and for the vast majority of 
interesting observables is) of the form :
\begin{equation}
V(p,k) = d_l \left (\frac{k_t}{Q}\right)^{a_l} e^{-b_l \eta} g_l(\phi)
\end{equation}
where $k_t$, $\eta$ and $\phi$ denote the transverse momentum, rapidity and azimuth wrt the hard (emitting) leg $l$ and $p$ the set of 
Born (hard) momenta. 
The parameters $a_l$, $b_l$ and $\phi$ can be numerically obtained.

The next step is to determine whether the observable
satisfies certain conditions, not previously available in the literature, 
which guarantee the validity of standard approximations that underly resummation methods. 
The most important condition is that of {\it{recursive}} infrared and collinear (IRC) safety which can be formulated as:
\begin{equation}
\left [ 
{\mathrm{lim}}_{\epsilon \to 0},{\mathrm{lim}}_{\epsilon' \to 0} \right ]
\frac{1}{\epsilon} V\left ( p,\epsilon k_1, \epsilon' \epsilon k_2,...\right )=0.
\end{equation}
In other words one probes the observable with two soft emissions (more precisely the commutator on the LHS of the above) 
rather than a single one (as for usual IRC safety) 
and checks that the scaling properties of the observable remain unaltered when one introduces the second soft emission. 
This condition being satisfied allows the exponentiation of the single gluon calculation (to NLL accuracy for global observables) 
with gluon branching merely reconstructing 
the running coupling in the exponent.

The final step is to insert the parameters determined above into an analytically computed master formula whose form depends on the number of hard legs in 
the Born configuration (e.g $n=4$ for hadron--hadron dijet event shapes). 
The final results are numerically determined by a Monte Carlo algorithm.
Results were presented (see \cite{salamproc} for the transverse thrust distribution at the Tevatron collider, 
as an example of the power of the 
automated resummation method.

LEP data continue to be an important source of information on event shapes 
as they are extracted in the cleanest possible environment for such studies.
The LEP QCD working group has made important progress in quantifying and understanding better the 
theoretical uncertainties associated with event shape predictions. For details on recent progress made at this meeting see \cite{fordproc}.

\input{sum2004}

\section{Future measurements}

With the LHC around the corner theoretical tools have to be improved 
to meet the challenge of making new discoveries in a very complex 
hadronic environment. Amongst the most prominent tools that will be 
employed in such searches for new physics will be Monte Carlo event generators.

Progress and developments concerning the new event generator SHERPA were presented in this conference \cite{Kraussproc}. This is an event generator that uses the matrix element (ME) package AMEGIC++ along with its own parton shower (PS) model and the CKKW procedure to merge the ME and PS. It also has an interface to the PYTHIA string fragmentation model for hadronization. 
Results from SHERPA for the jet $p_t$ distribution in 
Z+jet production at the Tevatron were shown, which indicated that the CKKW procedure, for such spectra, 
gave very similar results to those obtained from full NLO calculations.

There was also a presentation on the THEPEG \cite{Lonproc} 
which  is a general and modular C++ framework for implementing event generator models. Both PYTHIA7 and HERWIG++ will be built on THEPEG which will unify aspects of the event generation and share the administrative overheads involved.
In the same talk progress was also reported on the merging of matrix elements and parton showers for the ARIADNE Monte Carlo. Results obtained in the case of 
the $p_t$ distribution in W+jet production at the Tevatron indicated that the matching procedure worked extremely well in the incoming quark channel and less well for the incoming gluon channel with a visible cut-off artifact. 
Work is in progress in this regard.

\begin{figure}[!thb]
\begin{center}
\centerline{\epsfxsize=2.5in\epsfbox{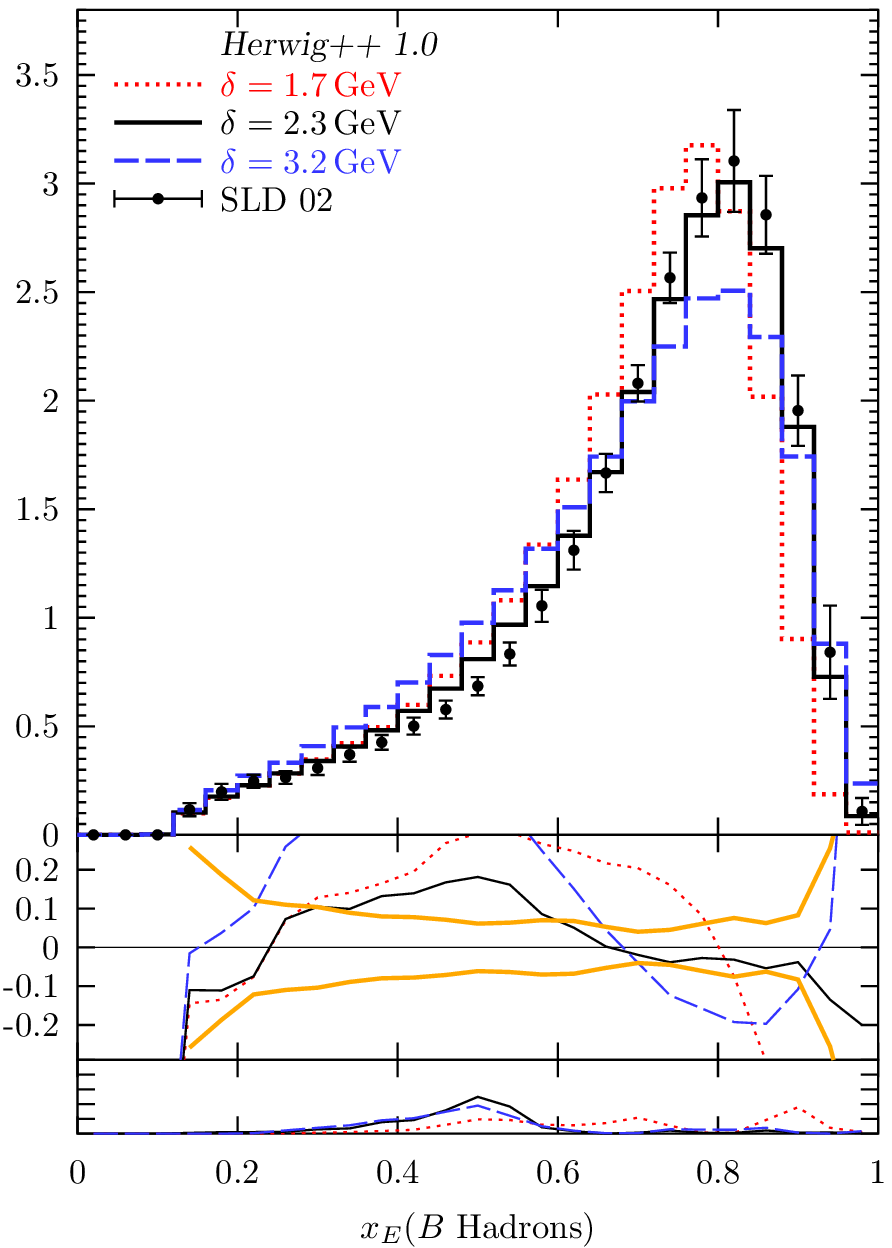}}
\end{center}
{\small Figure 10.
B fragmentation function from HERWIG++ compared to data for different values of the cut-off parameter $\delta$.}
\vspace{0.5cm}
\end{figure}

Results for event shapes, jet rates, heavy quark fragmentation functions etc.
were also displayed from the new event generator HERWIG++ \cite{Giesproc}, 
which was shown to perform as well\footnote{Thus far only $e^{+}e^{-}$ processes are implemented}(and in some cases better) than the fortran HERWIG, with precision tuning to LEP data still to be carried out.
It was emphasised that the C++ version is not merely a rewrite of the fortran HERWIG 
but there are some changes in the physics models. 
The most significant is perhaps the use 
of a new evolution variable $\tilde{q}$ 
inspired by the heavy quark splitting function :
\begin{equation}
P_{gq}(z,q^2,m^2) = \frac{C_F}{1-z} \left [1+z^2-\frac{2m^2}{z \tilde{q}^2} \right ]
\end{equation} 
Using $\tilde{q}$ as the evolution variable ensures a better coverage of the 
soft gluon phase space for emission off a heavy quark, 
in particular the description of the collinear (dead-cone) region. Evidence for the success of the new evolution 
can be seen in the excellent description of the B fragmentation function by 
varying just the parton shower parameters (see Fig.~10).
There is also an improved cluster hadronization model employed in HERWIG++.
Extensions to include $ep$ 
and especially hadron-hadron processes are eagerly awaited.

\section{Summary}
A selected sample of presentations given at the DIS04 workshop (working group C)
discussed different areas of hadronic final states phenomena has been summarized.
The detailed comparisons of the experimental data with the latest theoretical
developments show good agreement with QCD.   
In spite of all these successes of QCD theory, research in this direction is far from complete. 
We hope that the increase of the collected luminosity at HERA and Tevatron colliders,
together with future experiments at LHC and a linear collider,  
as well as  further progress in theoretical QCD calculations, will
lead to new exciting results which would open up new areas  where QCD theory can be confronted
with greater challenges.

\section*{Acknowledgments} 
We thank all DIS04 participants
for their talks as well as those who contributed to the lively discussions following
the presentations. 
This work was
supported in part by the High Energy Foundation and the World
Science Agency. 
We also thank all the organisers for the very exciting workshop and a perfect organisation.

\end{document}

%% file: intro.tex
\section{Introduction}
One of the still open challenges in modern physics is to understand more completely the strong
interaction. This is particularly important in light of the fact that 
many of the most important future discoveries in particle physics are expected to be made at the LHC, a machine that will collide strongly interacting particles at extremely high energies.
 
Hadronic final states produced in high-energy particle collisions have
traditionally been used as a testing ground for the theory of the 
strong 
interactions, quantum chromodynamics (QCD). Within our working group, 
results on hadronic final states from $ep$, $p\bar{p}$, $NN$ and $e^+e^-$ 
induced final states were
reported by 27 experimentalists and 11 theorists. The working group contained a vibrant atmosphere and was the forum for 
much useful discussion. 

The breadth of topics included precision tests of perturbative QCD with jet and leading particles. Perturbative QCD is in 
good shape in the regions in which it is applicable. However, results from H1 and ZEUS on forward jets and 
azimuthal decorrelations between jets suggest that the DGLAP evolution approach may be failing and that calculations including small Bjorken $x$ physics (BFKL/CCFM resummations) may be required. New measurements and theoretical predictions for event shapes, including new 
resummed calculations, have given improved consistency in the extraction of 
$\alpha_s$ and the non-perturbative parameter $\bar{\alpha}_0$. 
Amongst the main highlights 
of our working group were studies of exotic pentaquark states. 
In this working group, 
several experiments have reported their results on the search for 
pentaquark states 
with strangeness.

%% file: jetshigh.tex
\section{Jets and high $P_T$ phenomena}
\subsection{Progress in experimental tests of perturbative QCD}

The production of jets and leading particles provides a clean signature 
for probing perturbative QCD processes. At this workshop a number of measurements were 
presented from $ep$, $p\bar{p}$ and $e^+e^¯$ colliders. 

Multi-jet production in DIS provides an important testing ground 
for QCD. At HERA, data are collected over a large range of the negative 
four-momentum transfer squared, $Q^2$, the Bjorken-$x$ variable and the 
transverse energy $E_T$ of the observed jets. HERA dijet data can be used 
to gain insight into the dynamics of the parton cascade in low $x$ 
interactions. 

\begin{figure}[!thb]
\begin{center}
\centerline{\epsfxsize=3.5in\epsfbox{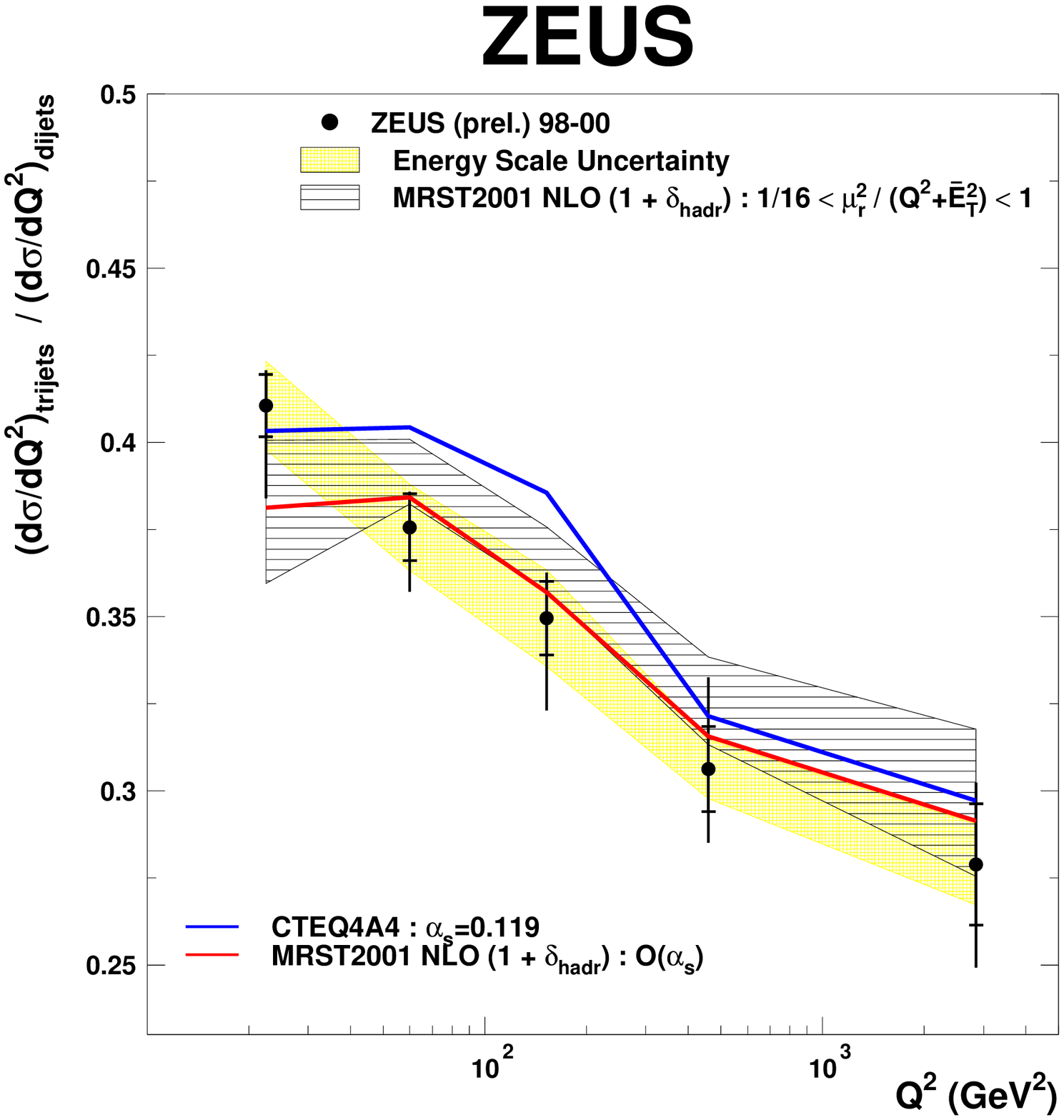}}
\caption[*]{The 3/2 jet ratio in DIS as measured by ZEUS \cite{ll}.
\label{figzeus32}
}
\end{center}
\end{figure}

ZEUS presented results on dijet and 3-jet production in DIS \cite{ll}. The 
data are 
described well by NLO calculations. The ratio of 3 and 2-jet production 
cross-sections is shown in Fig.~\ref{figzeus32}. In principle,  
experimental and theoretical uncertainties should be minimised by taking 
this ratio. However, the theoretical uncertainty owing to the choice of 
parton distribution function (pdf) is clearly large and this requires 
further investigation.

\begin{figure}[!thb]
\begin{center}
\centerline{\epsfxsize=3.5in\epsfbox{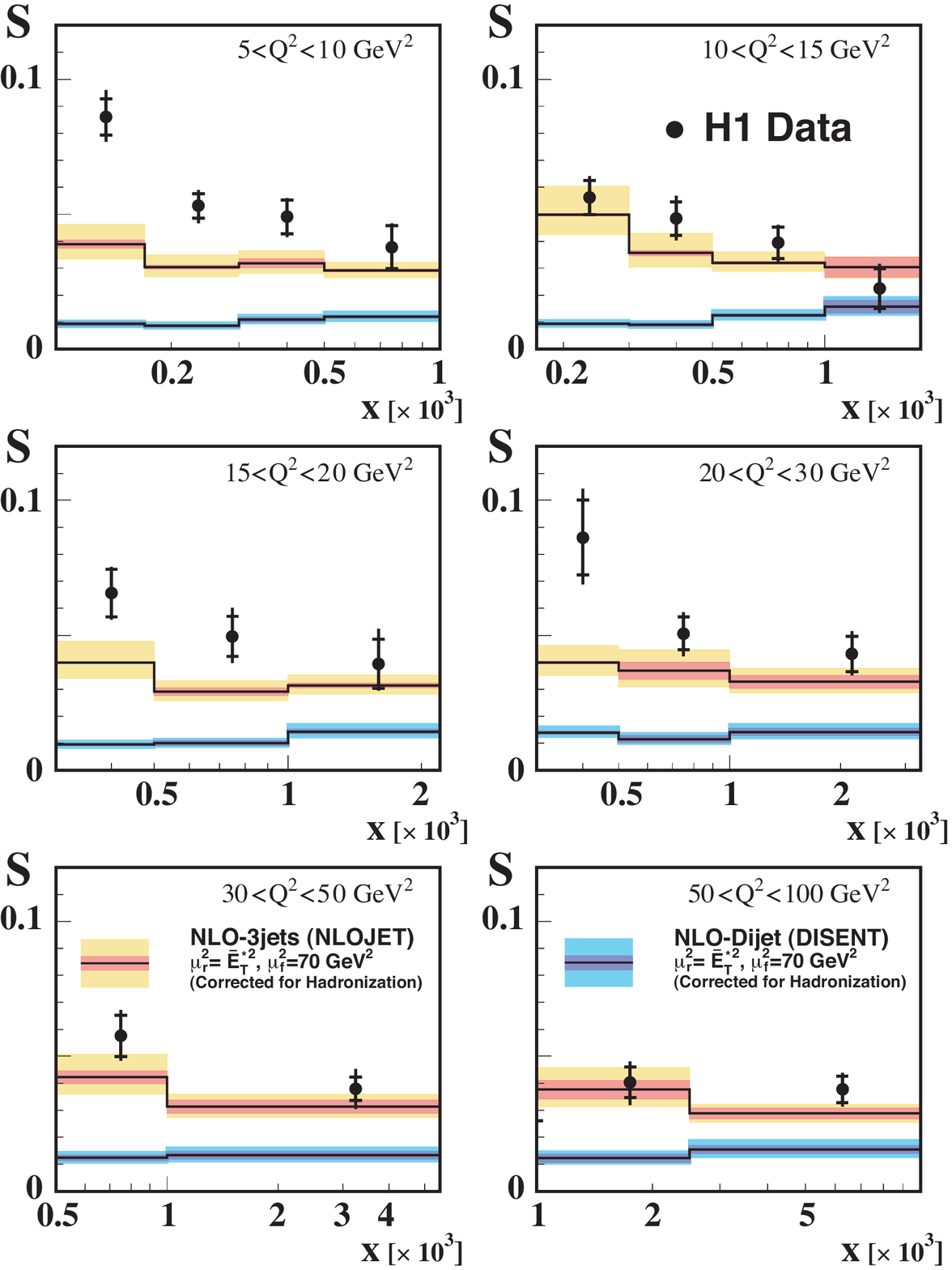}}
\caption[*]{The S distribution for dijet events, as measured by 
H1 \cite{rp}.
\label{figrp1}   
}
\end{center}
\end{figure}

Insight into low $x$ dynamics can be gained from inclusive dijet data by 
studying the behaviour of events with a small azimuthal separation between 
the two hardest jets as measured in the hadronic centre-of-mass system. 
Fig.~\ref{figrp1} shows measurements by H1 \cite{rp} of the $S$ 
distribution, i.e. the fraction of 
fraction of dijet events with an azimuthal separation of less than 
$120^\circ$ as a function of $x$ for different $Q^2$ regions. NLO 
calculations for 3 and 2-jet processes are shown. Although the 3-jet 
calculation comes closest, neither calculation is 
able to reproduce the data at 
the lowest $Q^2$ and  $x$ values. In this region, higher order, small $x$, 
QCD 
evolution effects 
are expected to become important and the validity of the DGLAP approach 
is in doubt..

\begin{figure}[!thb]
\begin{center}
\centerline{\epsfxsize=3.5in\epsfbox{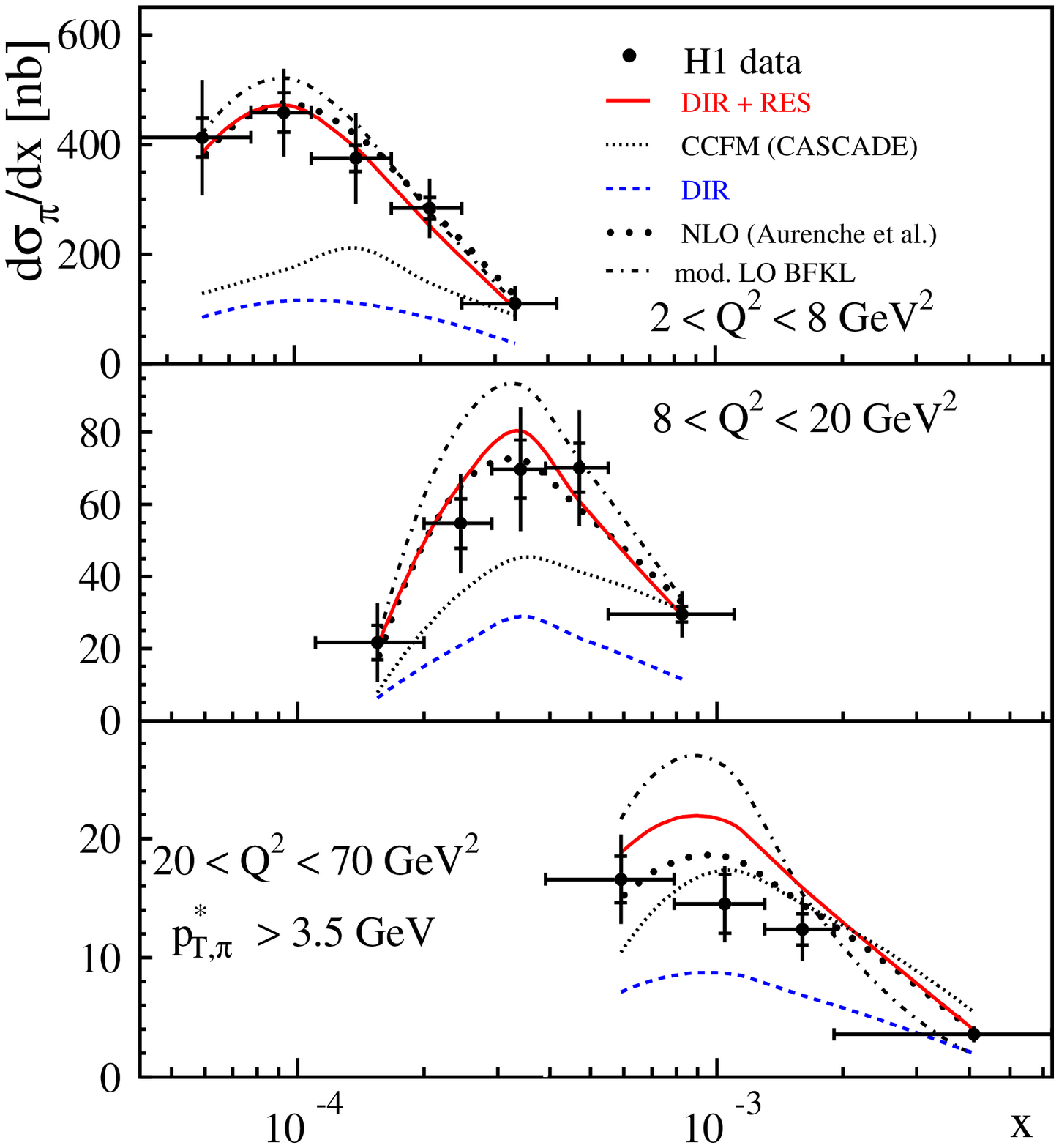}}
\caption[*]{The forward jet cross-section, as measured by H1 \cite{lg}. \label{figlg}
}
\end{center}
\end{figure}

A further way to experimentally probe QCD evolution is to study events 
containing high momentum jets \cite{ak} and particles \cite{lg} in the forward 
region i.e. next to the proton direction. Fig.~\ref{figlg} shows the 
cross-section for forward $\pi^\circ$-mesons produced at high transverse 
momentum as function of $x$. The data rise towards low $x$ and a QCD prediction based on DGLAP evolution for proton structure (labelled DIR) fails 
to describe the data. Approaches based on a resolved virtual photon 
picture (labelled DIR+RES) and on CCFM evolution describe the data better. 
Analytical calculations based on the BFKL evolution equation also give a 
reasonable description of the data. 

\begin{figure}[!thb]
\begin{center}
\centerline{\epsfxsize=3.5in\epsfbox{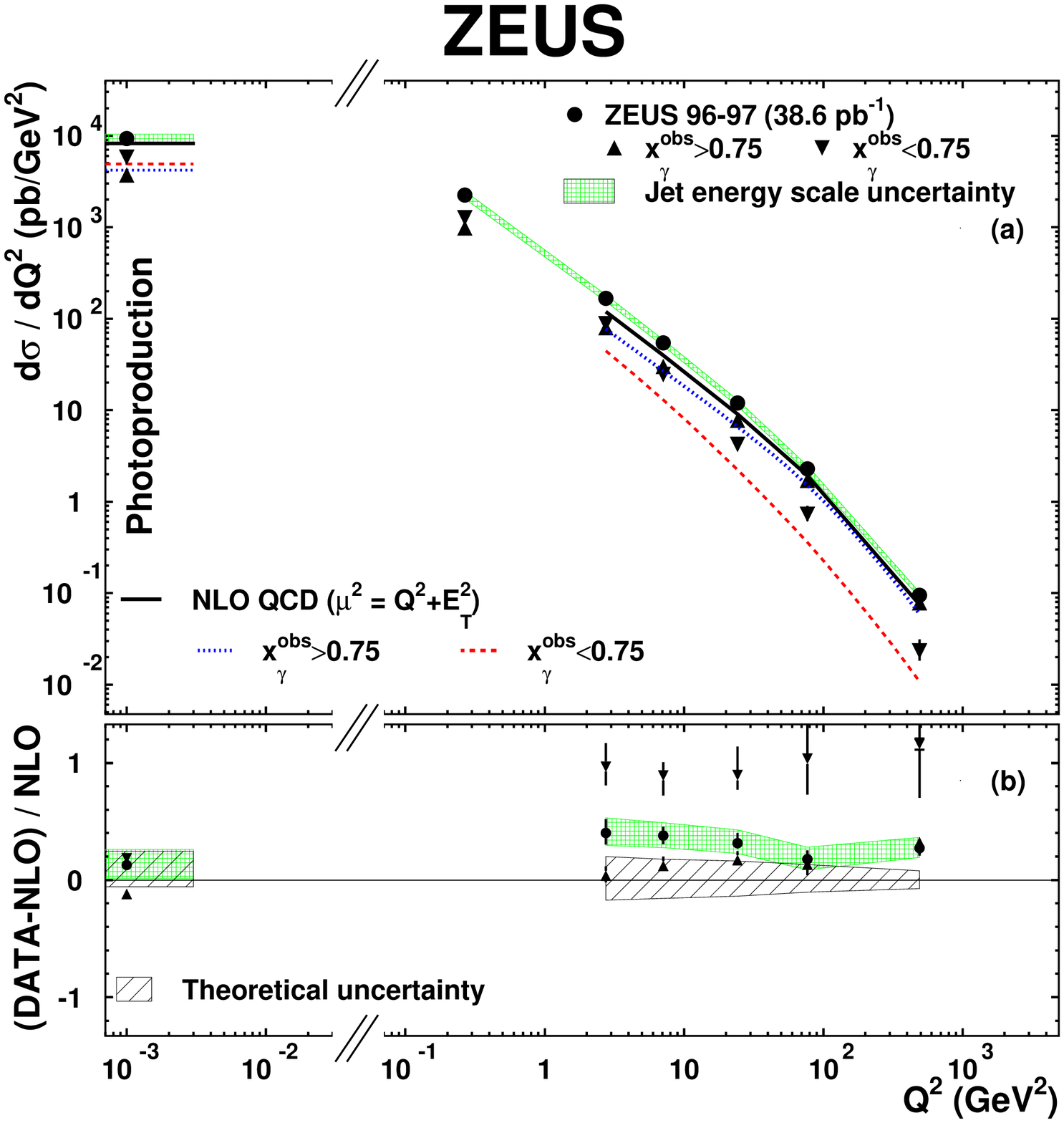}}
\caption[*]{The differential dijet cross-section, as measured by 
ZEUS.\label{figml}
}
\end{center}
\end{figure}

The structure of the photon at low $Q^2$ has been investigated by both the 
ZEUS \cite{ml} and H1 \cite{kse} experiments. Fig.~\ref{figml} shows the 
dijet 
cross-section as a function of $Q^2$ for different regions of $x_\gamma$, 
the variable used to determine the fraction of the photon's momentum 
taking part in the hard sub-process. The data are described by the NLO 
calculations for $x_\gamma >0.75$ although for $x_\gamma < 0.75$ the NLO 
prediction significantly underestimates the cross-section. 
This could be due to effects arising from the structure of the photon in 
this region. 

\begin{figure}[!thb]
\begin{center}
\centerline{\epsfxsize=3.5in\epsfbox{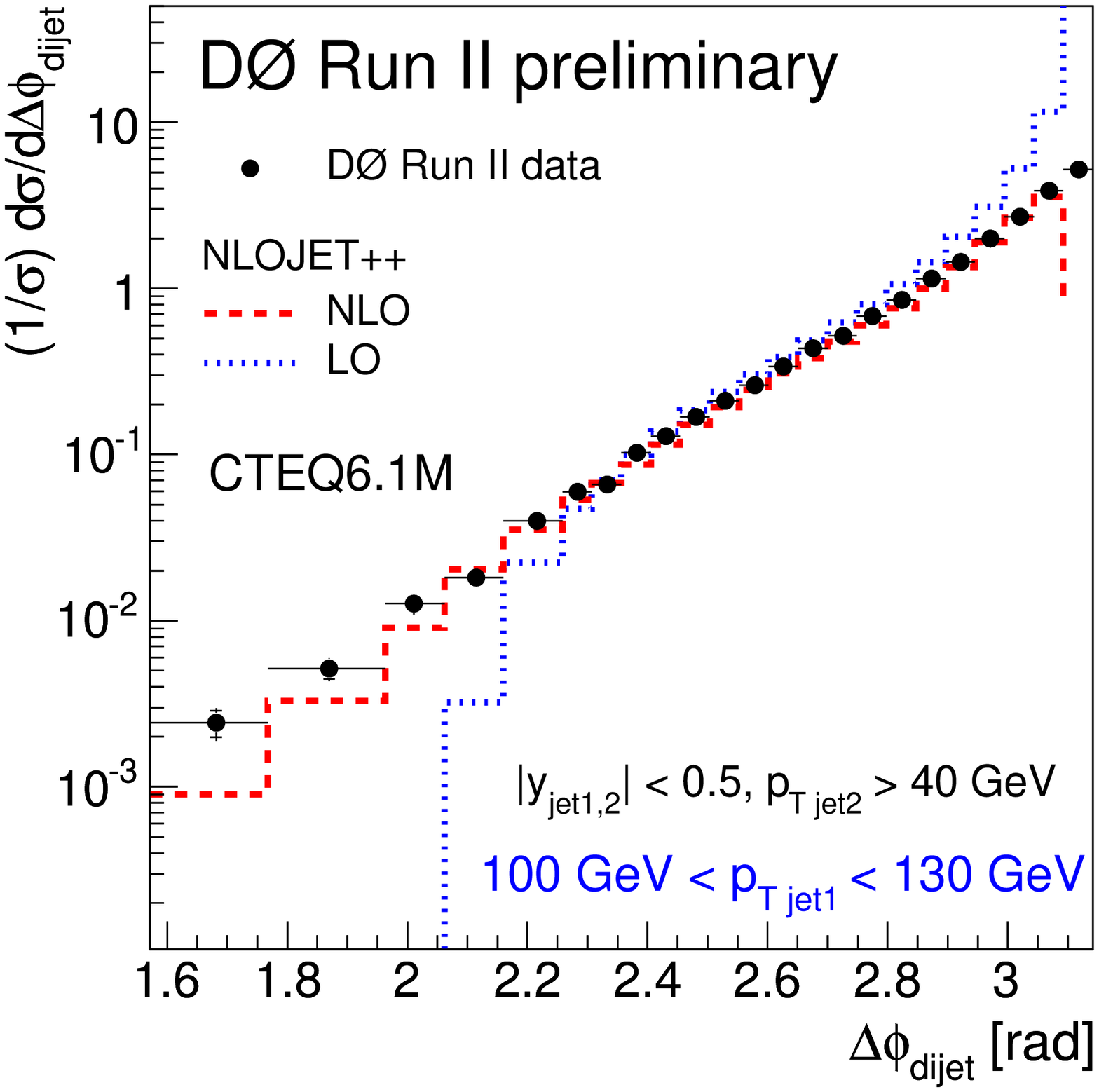}}
\caption[*]{The azimuthal decorrelation between dijets, as measured by D0.
\label{figd0}   
}
\end{center}
\end{figure}

Results from the CDF \cite{cdf}  and D0 \cite{d01,d02} showed encouraging 
progress in analysing Run II data. The increased centre of mass energy to 
1.96 GeV makes available a larger cross-section at the high values of jet 
transverse momentum. However, precision is severely limited due to 
calorimeter energy scale 
uncertainties for both experiments. The NLO predictions describe the data 
well although sensitivity is inhibited by the large 
energy scale uncertainty. In order to reduce the systematic effects from 
this source, D0 also presented results on the angular separation between 
the dijets, a variable which is highly sensitive to the higher order 
emissions. This is shown in Fig.~\ref{figd0}. NLO calculations provide 
an excellent description except for the extremely large (small) regions 
of $\Delta \Phi^*$, where the available phase space   
for multi-jet emissions is limited.

\begin{figure}[!thb]
\begin{center}
\centerline{\epsfxsize=4.5in\epsfbox{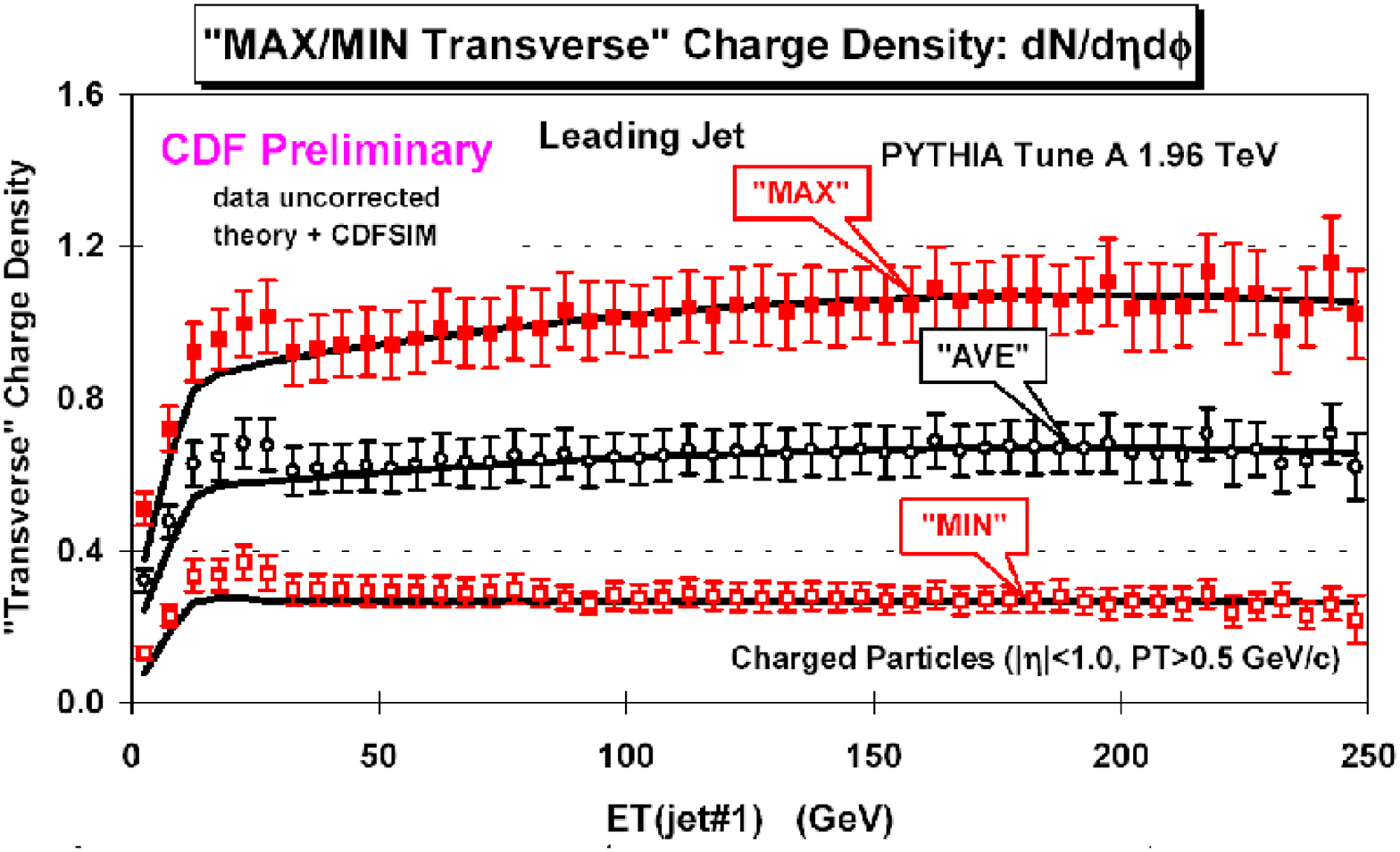}}
\caption[*]{The charged particle density for different selected regions of 
phase in dijet events, as measured by CDF \cite{cdfmb}.\label{cdfmb} }
\end{center}
\end{figure}

In collisions between objects with extended structure, the final state of 
hard parton-parton scattering is characterised by high transverse momentum 
jets but also contains a number of soft particles which are 
collectively referred to as the ``underlying event". This has been studied 
by CDF \cite{cdfmb} in which back-to-back dijet events are analysed in order 
to isolate the $\eta - \phi$ regions which are sensitive to the underlying 
event and to analyse its properties. Fig.~\ref{cdfmb} shows the charged 
particle density in the "MIN", "MAX" and average transverse region of 
leading jet events. On an event-by-event basis the "MAX" and "MIN" regions 
are defined to be the regions containing the largest and smallest, 
respectively, number of charged particles.  The density increases with the 
leading jet's transverse momentum in the "MAX" region and decrease in the 
"MIN" region. The complete results of this study suggest that there is a 
jet structure in the underlying event even when initial and final state 
radiation is strongly suppressed. The techniques and results of this work 
will be of enormous benefit at the LHC.

%% file: jetshighth.tex
\subsection{Theoretical developments}

Dijet data from H1 and ZEUS has been a significant source of information 
on quantities such as the pdfs. In such studies it has been commonly 
accepted that there is a need to impose 
asymmetric cuts on the transverse energy of the two highest $E_t$ jets to avoid problems of infrared 
enhancements. This involves cutting out a significant fraction of events since in fact the measured 
dijet total rate is largest when one allows symmetric cuts to be placed. 
Moreover, there is accurate data in the region of symmetric cuts and it is only the NLO QCD theoretical 
calculation that breaks down in this region. An improved theoretical estimate can be obtained by resumming large logarithms in the $E_t$ difference, $\Delta$,\footnote{More precisely it is the difference in $E_t$ cut values.} 
between the two highest $E_t$ jets \cite{banproc}.This allows one to use the theoretical calculation in the entire range where there is data. Moreover it allows one to test the ideas behind multijet resummations (in a problem less contaminated by power corrections than several event shapes) by comparing the next-to--leading log (NLL) 
resummed theory to the experimental data. In the absence, thus far, of 
any detailed comparison of theory and data for resummations with more than two hard  jets one would imagine that the calculation presented in \cite{banproc} provides a good opportunity for the same. 
Such multijet configurations will be especially important in the context of hadron colliders such as the LHC.
As far as results are concerned 
\begin{figure}[!thb]
\begin{center}
\centerline{\epsfxsize=3.5in\epsfbox{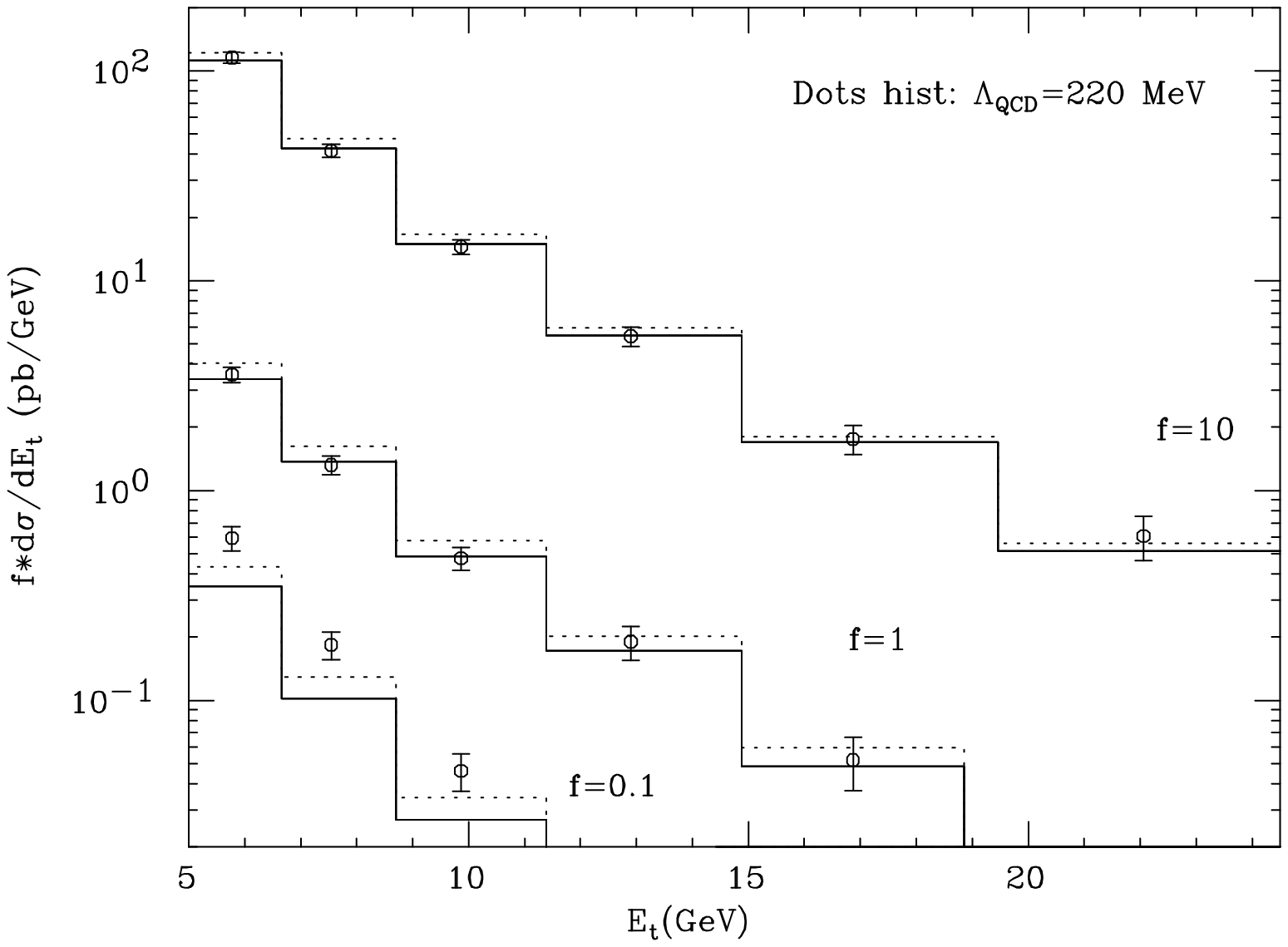}}
\caption[*]{Comparison between theory \cite{bertproc} and OPAL data for dijet production in $\gamma \gamma$ collisions in three 
different $x_\gamma$ regions. \label{figc}
}
\end{center}
\end{figure}
it was shown that after resummation the slope of the total rate $\sigma'(\Delta)$ was a linear function of $\Delta$ rather than a (double) logarithmically divergent one as indicated by NLO fixed-order computations. As a consequence the total rate itself rises monotonically with decreasing $\Delta$ instead of turning over as indicated by NLO alone. Future extensions involve extending the present cone algorithm 
results to the inclusive $k_t$ algorithm and carrying out the matching to fixed-order estimates.

Jet and dijet production in photon-photon collisions is also a valuable source of information on the hadronic structure of the photon. While NLO calculations previously existed in the framework of the phase-space slicing method 
\cite{KKK}, results were presented here \cite{bertproc} for NLO computations employing the dipole subtraction method which is preferable, especially 
in the case of computing certain differential cross sections, 
from the point of view of numerical stability. 
The new method also allows a cross-check of the previous phase-space 
slicing results. 
At present the comparisons with OPAL data are very satisfactory for 
inclusive observables e.g $E_t$ spectra in single inclusive jet production 
and the direct and single resolved dominated contributions to dijet 
production see Fig.~\ref{figc}. However one notes a discrepancy for the double resolved dominated region, which is still outstanding.

Understanding the multiple collinear limit of QCD amplitudes is of great importance in several contexts. 
These include the development of subtraction 
terms for fixed higher order (NLO and beyond) computations, 
all order resummations, more accurate pdf evolution and improving the physics content of Monte Carlo event generators.
For example the one loop correction to  one to three collinear branchings is one of 
the ingredients in computing the full NNLO pdf evolution.
Results were shown based on a new method that implements collinear factorisation directly in colour space thus retaining explicitly the colour matrix structure and resulting in the derivation of $1\to m$ splitting amplitudes. The (divergent part of) one loop splitting matrix in the limit of $m$ collinear partons reads (where $s_{1...m}$ denotes $(p_1+p_2+...p_m)^2$ and the $p_i$ are collinear parton momenta:
\beeq
\!\!\! \sp^{(1) \,{\rm div.}}(p_1,\dots,p_m) &=& 
\frac{\Gamma(1+\epsilon) \,
\Gamma^2(1-\epsilon)}{\left(4\pi\right)^{-\epsilon}\Gamma(1-2\epsilon)}
\; \frac{1}{2}
\left\{ \frac{1}{\ep^2}
\sum_{i,j=1 (i \neq j)}^m \;{\bom T}_i \cdot {\bom T}_j
\left( \frac{-s_{ij} -i0}{\mu^2}\right)^{-\ep} \right. \nn \\
&&\left. + 
\left( \frac{-s_{1 \dots m} -i0}{\mu^2}\right)^{-\ep}
\left[ \frac{1}{\ep^2}
\sum_{i,j=1}^m \;{\bom T}_i \cdot {\bom T}_j
\;\left( 2 - \left( z_i \right)^{-\ep} -\left( z_j \right)^{-\ep} \right)
\right. \right.  \nn \\
\label{sp1div}
&&\left. \left. - \frac{1}{\ep} 
\left( \sum_{i=1}^m \left( \gamma_i - \ep {\tilde \gamma}_i^{\RS} \right)
- \left( \gamma_a - \ep {\tilde \gamma}_a^{\RS} 
\right) - \frac{m-1}{2} \left( \beta_0 - \ep {\tilde \beta}_0^{\RS}
\right) \right) \right]
\right\} \nn \\
&\times& \sp^{(0)}(p_1,\dots,p_m) \;\;,
\eeeq
which shows the infrared and ultraviolet $\epsilon$ pole structure.
In the above the colour charge of the collinear parton with momentum $p_i$
is denoted by ${\bom T}_i$, and from colour conservation one has 
$\sum_i {\bom T}_i \;\sp^{(0)} = \sp^{(0)} \;{\bom T}_a$ (${\bom T}_a$
is the colour charge of the parent parton in the collinear splitting).
The flavour coefficients $\gamma_i$ 
and $\beta_0$
are $\gamma_q=\gamma_{\bar q}=3C_F/2$
and $\gamma_g=\beta_0/2=(11C_A-2N_f)/6$. The flavour coefficients 
${\tilde \gamma}_i^{\RS}$ and ${\tilde \beta}_0^{\RS}$ are renormalisation scheme 
dependent 
. $\sp^{(0)}(p_1,\dots,p_m)$ is the corresponding tree level quantity. From the above equation and the tree level result one can compute one loop splitting functions for the $1\to m$ collinear branchings \cite{rodproc}.

%% file: sum2004.tex
\section{Particle production}
The confinement transition from partons to the hadrons observed by experiments is poorly understood.   
Thus particle production phenomena cannot be explained in a systematic way solely from perturbative QCD. 
Studies of identified hadrons, inclusive hadronic multiplicities and particle flows 
allow to conduct detailed and sophisticated studies of many aspects of sort limit of QCD. At the same time, 
at this workshop we have seen how such observable can be used to tackle complicated problems 
related to perturbative QCD effects. 

\subsection{Conventional states}

Identification of conventional baryons and mesons  have received much attention in this working group.
In particular, baryon-antibaryon production in two-photon processes in  $e^+e^-$ collisions
has stable interest at LEP community. Agreements with the
diquark and the handbag predictions were found in reasonable shape 
for the L3 studies \cite{l3bar},
while the expectation from the three-quark
model  was clearly below the data.

Several results on the measurements of (anti)deutrons \cite{phenix,h1deut} 
were presented by both PHENIX and H1
collaborations. 
In heavy ion collisions, the production of (anti)nuclei helps to understand the production size
of the system and amount of radial expansion when the coalescence model is used to explain the production
mechanism of (anti)deutrons \cite{phenix}.
In particularly, deuteron/antideuteron spectra at mid-rapidity probe 
the late stages of relativistic heavy ion collisions.

The measurements of (anti)deutrons  are clearly a significant challenge for experimentalists, 
since these states 
can easily be produced by secondary scattering processes. 
This problem is less severe for antideutrons which
were measured for the first time in $ep$ collisions by the H1 collaboration \cite{h1deut}.  
The production mechanism of antideutrons in $ep$ collisions is significantly less understood compared
to heavy-ion collisions. 
The result of this interesting study was in the observation that the production rate
of antideutrons in $ep$ is by  
order of magnitude smaller than in heavy-ion collisions.
Thus, in terms of the coalescence model, the size of the fireball 
at the thermal freeze out is expected to be much smaller for $ep$ processes than for heavy-ion collisions.      

\subsection{Pentaquarks}

This was the first DIS workshop at which the entire section was devoted to searches for pentaquarks,
five-quark 
baryons which recently attracted tremendous attention of theorists and experimentalists.  
Let us remind that recent results from low-energy fixed-target experiments \cite{fixed} gave evidence for the
existence of a new narrow baryon resonance, $\Theta^+$  with a mass of approximately
1530 MeV and positive strangeness. This state was seen in  
in the $K^+ n$ decay channel.
These results have triggered new interest in baryon spectroscopy since
this baryon is manifestly exotic; it cannot be composed of three
quarks, but may be explained as a bound state of five quarks, i.e. as
a pentaquark, $\Theta^+ = uudd\bar{s}$. A narrow baryonic resonance
close to the observed mass is predicted in the chiral soliton
model \cite{zp:a359:305}. The quantum numbers of this state also permit
decays to $\ksp$ and $\kspb$, which can experimentally be measured using 
tracking detectors \cite{ks}.

Two DESY experiments, HERMES \cite{hermes} and ZEUS \cite{zeus}, reported observations of a possible
pentaquark state in the decay mode $\ksppb$, see Fig.~9. The masses and widths agreed  rather  well
for both measurements. However, some discrepancies with the $K^+ n$ decay channel \cite{fixed} may exist. 
This ether could mean that the particle decaying to  $\ksppb$ is not the same state 
as that observed for the $K^+ n$ channel,
or, unlike the well reconstructed $\ksppb$ final state, measurements involving detection of neutrons 
using missing mass could suffer from a significant systematic
uncertainty which is not fully understood.     

\begin{center}
\vspace{0.5cm}
\begin{minipage}[c]{0.48\textwidth}
\includegraphics[width=4.0cm,angle=0]{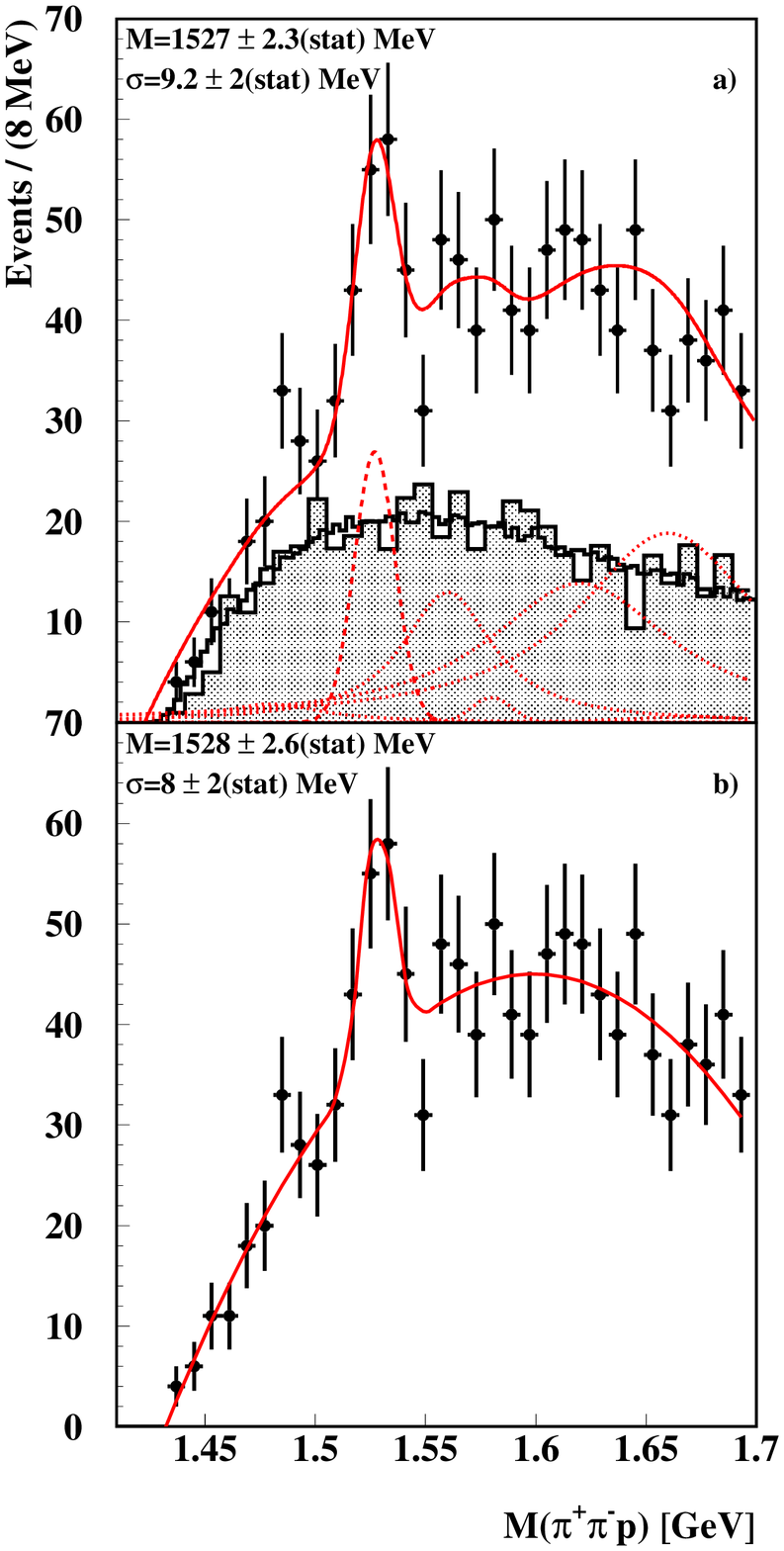}
\label{theta++}
\end{minipage}
%
\begin{minipage}[c]{0.48\textwidth}
\includegraphics[width=6.5cm,angle=0]{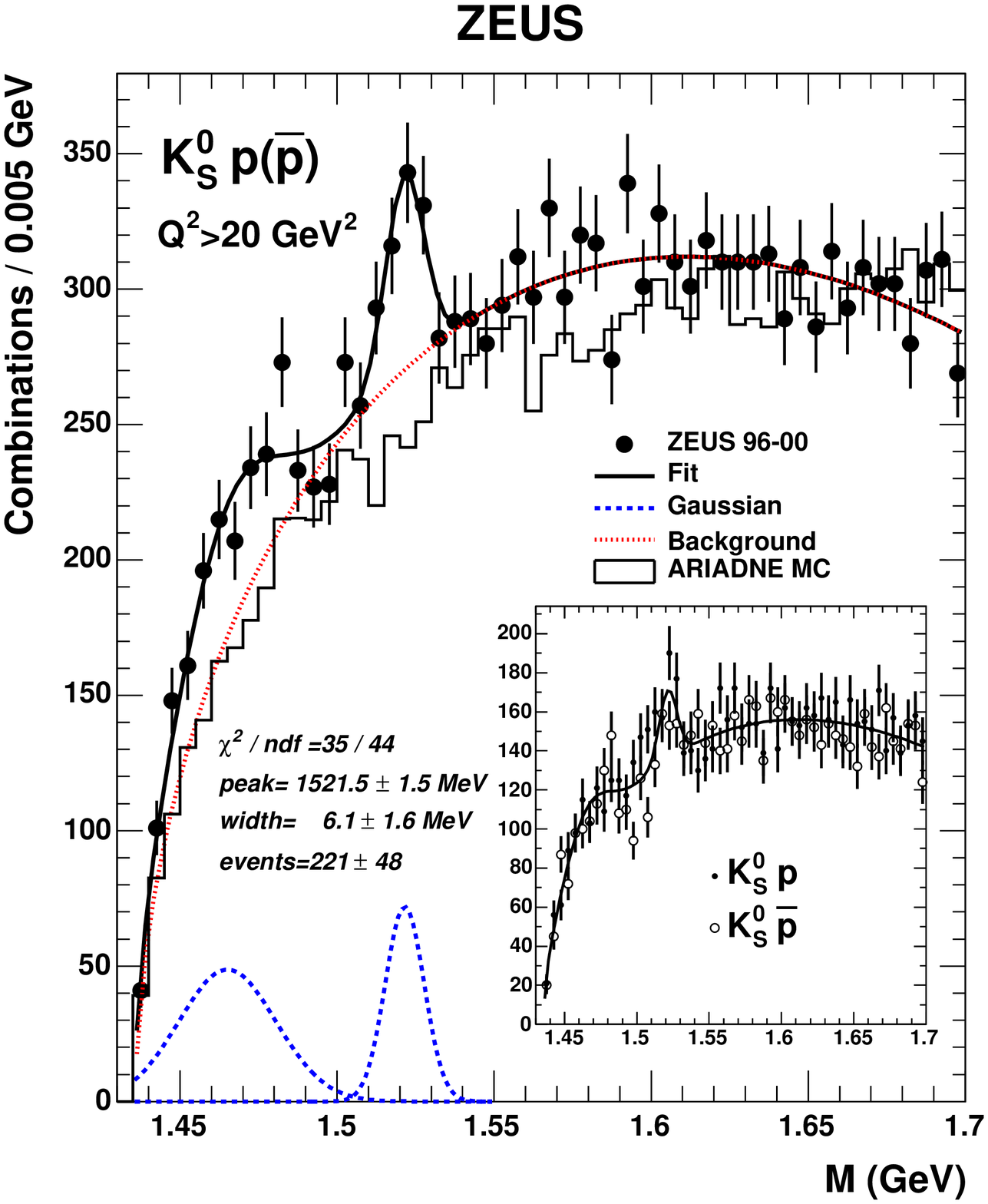}
\end{minipage}
\end{center}
{\small 
Figure 9: 
Invariant-mass spectra for the $\ksppb$ channel measured by HERMES \cite{hermes} and ZEUS \cite{zeus} experiments 
at DESY. The measurements were performed for photoproduction (HERMES) and DIS (ZEUS) data.}
\vspace{0.5cm}

ALEPH reported negative results on the  $\Theta^+$ search \cite{aleph}. Presently, other  LEP
experiments have also reported their negative results (see a recent review \cite{pen_rev}). 
Does this related to the fact that $e^+e^-$ experiments do not have a contribution from the proton
remnant, i.e. from the net baryon number in the initial state? 
At this conference there was a full concensus that 
this question requires further studies and, at present, no definite answer can be given.

Note that ZEUS is the first high-energy experiment which made the observation of $\Theta^+$ state in the central
fragmentation region dominated by fragmentation of quarks and gluons. 
One should mention that, unlike fixed-target
experiments, the contribution from the proton remnant is negligible in this region. Thus, if confirmed, this result
needs to be understood in terms of present hadronisation models. 

It is important to point out again that several high-energy 
experiments have reported their negative results (see references
in \cite{pen_rev}). Of course, each negative case should be considered very seriously.    
If the pentaquark does exist near $1520-1530$ MeV, possible
reasons of such ZEUS success may lay in a good tracking resolution for  $\ksppb$ combinations 
($\simeq 2.5$ MeV) and a relatively small combinatorial background for DIS at $Q^2>20$ GeV$^2$ (compared to some heavy-ion
and Tevatron experiments). Of course, differences in the production dynamics may also be important.   
One possible way to learn about instrumental and background
differences between different experiments would be to use the PDG $\Lambda_c$ baryon 
(which can also decay to $\ksppb$) as a reference.
Both HERMES and ZEUS used this state to calibrate the mass measurements near the 
$\ksppb$ production threshold.

The $\Theta^+$ lies at the apex of a hypothetical anti-decuplet of pentaquarks
with spin $1/2$. The baryonic states
$\Xi^{--}_{3/2}$ and $\Xi^{0}_{3/2}$ at the bottom of this
antidecuplet  are also manifestly exotic.
According to the predictions of  Diakonov et al. \cite{zp:a359:305},  the
members in the anti-decouplet with the isospin
quartet of $S=-2$ baryons should have a mass of
about $2070\mev$ and partial decay width into $\Xi\pi$ of about
$40\mev$. 
Recently, NA49 \cite{prl92:042003} at the CERN SPS
made observation
of exotic $\Xi^{--}_{3/2}$ member of the
$\Xi$ multiplet, with a mass of $1862\pm 2\mev$ and width below the
detector resolution of about $18\mev$.
At this conference, several experiments, CDF, ALEPH and ZEUS reported their negative results on the search
for $\Xi^{--}_{3/2}$ and $\Xi^{0}_{3/2}$.
Some experiments have extremely competitive data with a large number
of reconstructed $\Xi$ states used  for the searches of exotic $\Xi^{--}_{3/2}$ 
and $\Xi^{0}_{3/2}$.

\subsection{Multiplicities and particle flows}

The measurements of multihadronic  final state allow
detection of color coherence properties.
A number of interesting results were shown 
by DELPHI collaboration \cite{delphi_mul} using hadron multiplicities: a competitive value for the  
color-factor ratio was obtained, a clear signature for the destructive gluon interference was seen, and finally,
the ratio of gluon to quark multiplicities was  obtained and compared to the 3NLO and MLLA calculations. Both predictions
gave a reasonably good agreement to the data.    

Particle flows were also used by DELPHI \cite{delphi_color}  to study
color-reconnection effects between color
singlets which take place in $WW$ production.
This phenomenon contributes to enhancement or depletion of soft particle multiplicity in some phase-space regions.
The data might indicate that there is a small contribution from color reconnection effect,
but generally, the predictions are doing rather well without extra contributions. A  possible mass
shift in  the reconstructed $W$ mass is expected to be rather small, by order of magnitude smaller than that
expected in early days of the $WW$ studies.

Clearly, particle flows can be used to test certain details of hard QCD without using the jet algorithms, as long as we
understand contributions from soft physics to such observables.
There has been considerable progress in studying  
azimuthal asymmetry attributed to hard QCD by using the energy 
flow method as reported by ZEUS \cite{zeus_flow}. It was shown that this method has
several advantages compared to conventional multiplicity-based  and jet-based experimental techniques. The results
indicate a rather good agreement with the NLO calculations (corrected by Monte Carlo models to take into account
hadronisation effects).  

\subsection{Theoretical developments in particle production}
Results were also presented in this meeting, on NLO computations
for inclusive and two photon production at hadron colliders \cite{Guilletproc}.
Successful comparisons were shown of the NLO predictions from the program DIPHOX with 
data from CDF and D0 Run II data for quantities such as azimuthal angle, invariant mass and
the transverse momentum $p_t$ of the $\gamma$ pair in two photon production. In regions requiring resummation
(such as the region of small $p_t$ of the pair), as expected the comparison with NLO is less
meaningful. However it was clear that two photon production at the Tevatron is well 
understood theoretically while a need for more data was observed in the inclusive photon case.